\documentclass[11pt]{article}
\usepackage{graphicx,amsfonts,amssymb,amsmath}
\newtheorem{thm}{{\bf Theorem}}[section]
\newtheorem{lemma}[thm]{{\bf Lemma}}
\newtheorem{prop}[thm]{{\bf Proposition}}
\newtheorem{cor}[thm]{{\bf Corollary}}

\def\be{\begin{eqnarray}}
\def\ee{\end{eqnarray}}
\def\ben{\begin{eqnarray*}}
\def\een{\end{eqnarray*}}
\def\ba{\begin{array}}
\def\ea{\end{array}}
\def\bp{\noindent{\it Proof. }}
\def\ep{\noindent{\hfill \fbox{}}}

\def\wt{\widetilde}
\def\remark{\noindent{\bf Remark. }}
\def\definition{\noindent{\bf Definition}}
\def\pic{\rm Pic}

\def\ni{\noindent}
\def\vp{\varphi}
\def\al{\alpha}

\def\ponet{{\mathbb P}^1\times{\mathbb P}^1}
\def\mc{{\mathbb C}}
\def\mz{{\mathbb Z}}

\newcommand{\ol}[1]{\overline{#1}}
\newcommand{\mapright}[1]{%
   \smash{\mathop{%
   \hbox to 1cm{\rightarrowfill}}\limits^{#1}}}
\newcommand{\mapleft}[1]{%
   \smash{\mathop{%
   \hbox to 1cm{\leftarrowfill}}\limits^{#1}}}
\newcommand{\maplleft}[2]{%
   \smash{\mathop{%
   \hbox to 1cm{\leftarrowfill}}\limits_{#1}^{#2}}}

\begin{document}
\title{The extended Weyl group $\widetilde{W}(D_5^{(1)})$ as an extension of 
KNY's birational representation of $\widetilde{W}(A_1^{(1)}\times A_3^{(1)})$}
\author{Tomoyuki Takenawa}
\date{}
\maketitle
\begin{center}
{Graduate School of Mathematical Sciences, University of
Tokyo, Komaba 3-8-1, Meguro-ku, Tokyo 153-8914, Japan}\\
\end{center}

\begin{abstract}
We study the birational representation of $\wt{W}(A_1^{(1)}\times A_3^{(1)})$
proposed by Kajiwara-Noumi-Yamada (KNY) in the case of $m=2$ and $n=4$.
It is shown that the equation 
can be lifted to an automorphism of a family of $A_3^{(1)}$ surfaces 
and therefore the group of Cremona isometries is $\wt{W}(D_5^{(1)})$
($\supset \wt{W}(A_1^{(1)}\times A_3^{(1)})$).
The equation can be decomposed into two mappings 
which are conjugate to the $q$-$P_{VI}$ equation.
It is also shown 
that the subgroup of Cremona isometries which commute with the original 
translation is isomorphic to 
$\mz \times \wt{W}(A_3^{(1)}) \times \wt{W}(A_1^{(1)})$.
\end{abstract}


\section{Introduction}

Since the singularity confinement criterion was introduced
as a discrete analogue of the Painlev\'e test \cite{grp},
many discrete analogues of Painlev\'e equations have been proposed
and extensively studied \cite{js,rgh}.
Discrete Painlev\'e equations have been considered as 
2-dimensional non-autonomous
birational dynamical systems which satisfy this criterion and which
have limiting procedures
to the (continuous) Painlev\'e equations.
In recent years it was shown by Sakai that all
of these (from the point of view of symmetries) are
obtained by studying rational surfaces in connection with extended
affine Weyl groups \cite{sakai}. 

On the other hand, recently Kajiwara {\it et al} (KNY) 
\cite{kny2} have proposed a 
birational representation of the extended Weyl groups
$\wt{W}(A^{(1)}_{m-1}\times A^{(1)}_{n-1})$
on the field of rational functions ${\mathbb C}(x_{ij})$,
which is expected to provide higher order discrete 
Painlev\'e equations
(however, this representation is
not always faithful, for example it is not faithful
in the case where $m$ or $n$ equals $1$ and
in the case of $m=n=2$).
In the case of $m=2$ and $n=3,4$, the actions of the translations can be 
considered to be 2-dimensional non-autonomous discrete dynamical systems
and therefore to correspond to discrete Painlev\'e equations. 
Special solutions and 
some properties of these equations have been studied by several authors 
\cite{kny1,masuda}.   
In the case of $m=2$ and $n=4$, the action of the translation was thought
to be a symmetric form of the $q$-discrete analogue of Painlev\'e V equation 
($q$-$P_V$). 
However, the symmetry $\wt{W}(A^{(1)}_1 
\times A^{(1)}_3)$ does not coincides with any symmetry of 
discrete Painlev\'e equations in Sakai's list, 
(in the case of $m=2$ and $n=3$,
it coincides with an equation, which is associated with a family of 
$A_3^{(1)}$ 
surfaces and whose symmetry is  $\wt{W}(A^{(1)}_1 \times A^{(1)}_2)$, 
in Sakai's list). 
So it is natural to suspect
that the symmetry might be a subgroup of a larger group associated with 
some family of rational surfaces.

In this paper we show that in the case of $m=2$ and $n=4$ the action
of the translation can be lifted to an automorphism of a family 
of rational surfaces of the type $A_3^{(1)}$,
i.e. surfaces such that the type of the configuration of irreducible 
components of their anti-canonical divisors is $A_3^{(1)}$,  
and therefore that the group of these automorphisms is $\wt{W}(D_5^{(1)})$
(hence it is not $q$-$P_V$ by Sakai's classification).
The action can be decomposed into two mappings 
which are conjugate to the $q$-$P_{VI}$ equation.
It is also shown 
that the subgroup of automorphisms which commute with the original 
translation is isomorphic to $\mz \times 
\wt{W}(A_3^{(1)}) \times \wt{W}(A_1^{(1)})$.


\section{ Birational representation of 
$\wt{W}(A^{(1)}_{m-1} \times A^{(1)}_{n-1})$}

The birational representation of 
$\wt{W}(A^{(1)}_{m-1} \times A^{(1)}_{n-1})$
on ${\mathbb C}(x_{i,j})$ proposed by Kajiwara {\it et al} (KNY) \cite{kny2}
is an action on ${\mathbb C}(x_{i,j})$
($i=0,1,\ldots,m-1$, $j=0,1,\ldots,n-1$
and the indices $i,j$ are considered
in modulo $m{\mathbb Z}, n{\mathbb Z}$ respectively) defined as follows.

We write the elements of the Weyl group corresponding to the simple roots as 
$$r_i \in W(A^{(1)}_{m-1}),~s_j \in W(A^{(1)}_{n-1})$$ 
and the elements corresponding to the rotations of the Dynkin diagrams as
$$\pi \in {\rm Aut}({\rm Dynkin}(A^{(1)}_{m-1})),~ 
\rho\in {\rm Aut}({\rm Dynkin}(A^{(1)}_{n-1})).$$
The action of these elements on ${\mathbb C}(x_{i,j})$ are defined as
\ben
&&r_i(x_{ij})=x_{i+1,j} 
\frac{P_{i,j-1}}{P_{ij}}, \quad
r_i(x_{i+1,j})=x_{ij}
\frac{P_{ij}}{P_{i,j-1}}, \quad
r_k(x_{ij})=x_{ij}, \quad (k \neq i,i-1),\\[2mm]
&&s_j(x_{ij})=x_{i,j+1} 
\frac{Q_{i-1,j}}{Q_{ij}}, \quad
s_j(x_{i,j+1})=x_{ij}
\frac{Q_{ij}}{Q_{i-1,j}}, \quad
s_k(x_{ij})=x_{ij}, \quad (k \neq j,j-1),\\[2mm]
&&\pi(x_{ij})=x_{i+1,j}, \quad \rho(x_{i,j})=x_{i,j+1}, 
\een
where 
\ben
P_{ij}=\sum_{a=0}^{n-1} \left( 
\prod_{k=0}^{a-1} x_{i,j+k+1} 
\prod_{k=a+1}^{n-1} x_{i+1,j+k+1} \right), &&
Q_{ij}=\sum_{a=0}^{m-1} \left( 
\prod_{k=0}^{a-1} x_{i+k+1,j} 
\prod_{k=a+1}^{m-1} x_{i+k+1,j+1} \right).
\een
For example in the $(m,n)=(2,4)$ case,
$$P_{00}=x_{1,2} x_{1,3} x_{1,0}+x_{0,1} x_{1,3} x_{1,0}
  +x_{0,1} x_{0,2} x_{1,0}+x_{0,1} x_{0,2} x_{0,3}$$
and $Q_{00}=x_{0,1}+x_{1,0}$. 
It was shown by KNY that this action is a representation
of $\wt{W}(A^{(1)}_{m-1} \times A^{(1)}_{n-1})$ as automorphisms 
of the field ${\mathbb C}(x_{i,j})$. But it is still an open problem 
when this representation is faithful. In the case of $m=2$ and $n=3,4$, 
one can see it is faithful 
by considering the actions on the root systems which we discuss later.

In the case of $(m,n)=(2,4)$, the variable transformation:
\ben
\left( \frac{x_{0,j}x_{1,j}}{x_{0,j+1}x_{1,j+1}}\right)^{1/2}=a_j,&&
\left( \frac{x_{0,j} x_{0,j+1}}{x_{1,j} x_{1,j+1}}\right)^{1/2}=f_j~,
\een
reduces the actions of $r_0,r_1,s_0,s_1,s_2,s_3,\pi,\rho$ to
\ben
&&r_0(a_i)=a_i,~~\\
&&r_0(f_i)=\frac{1}{a_{i}a_{i+1}f_{i+1}}
\frac{1+a_{i}f_{i}+a_{i}a_{i+1}f_{i}f_{i+1}+
                a_{i}a_{i+1}a_{i+2}f_{i}f_{i+1}f_{i+2}}
{1+a_{i+2}f_{i+2}+a_{i+2}a_{i+3}f_{i+2}f_{i+3}+
                a_{i+2}a_{i+3}a_{i+4}f_{i+2}f_{i+3}f_{i+4}}\\
&&\pi(a_i)=a_i, ~~~~\pi(f_i)=\frac{1}{f_i}\\
&&r_1=\pi \circ r_0 \circ \pi\\
&&s_i(a_j)=a_ja_i^{-c_{i,j}},~~~~
s_i(f_j)=f_j \left(\frac{a_i+f_i}{1+a_if_i}\right)^{u_{i,j}}\\
&&\rho(a_i)=a_{i+1},~~~~\rho(f_i)=f_{i+1}
\een
where $c_{i,j}$ and $u_{i,j}$ are 
\ben
(c_{i,j})=
 \left(
  \begin{array}{rrrr}
    2 & -1 &  0 & -1 \\
   -1 &  2 & -1 &  0 \\
    0 & -1 &  2 & -1 \\
   -1 &  0 & -1 &  2
  \end{array}
 \right),   
&&(u_{i,j})=
 \left(
  \begin{array}{rrrr}
    0 &  1 &  0 & -1 \\
   -1 &  0 &  1 &  0 \\
    0 & -1 &  0 &  1 \\
    1 &  0 & -1 &  0
  \end{array}
 \right).
\een

\remark
By the variable transformation we have $a_0 a_1 a_2 a_3=1$,
but if we remove this constraint and set $a_0 a_1 a_2 a_3=q^{-1}$, 
the actions also generate 
$\wt{W}(A^{(1)}_{m-1} \times A^{(1)}_{n-1})$.\\

The element $\pi \circ r_0$ is a translation of $A_1^{(1)}$
and provides a discrete dynamical system:
\ben
\begin{array}{c}
 \smallskip
 \displaystyle 
 \bar{a}_0=a_0, \quad \bar{a}_1=a_1, \quad \bar{a}_2=a_2, \quad \bar{a}_3=a_3,
 \\[2mm]
 \smallskip
 \displaystyle 
 \bar{f}_0=a_0 a_1 f_1 
           \frac{1+a_2 f_2+a_2 a_3 f_2 f_3 +a_2 a_3 a_0 f_2 f_3 f_0}
                {1+a_0 f_0+a_0 a_1 f_0 f_1 +a_0 a_1 a_2 f_0 f_1 f_2}, \\[2mm] 
 \smallskip
 \displaystyle 
 \bar{f}_1=a_1 a_2 f_2 
           \frac{1+a_3 f_3+a_3 a_0 f_3 f_0 +a_3 a_0 a_1 f_3 f_0 f_1}
                {1+a_1 f_1+a_1 a_2 f_1 f_2 +a_1 a_2 a_3 f_1 f_2 f_3}, \\[2mm]
 \smallskip
 \displaystyle 
 \bar{f}_2=a_2 a_3 f_3 
           \frac{1+a_0 f_0+a_0 a_1 f_0 f_1 +a_0 a_1 a_2 f_0 f_1 f_2}
                {1+a_2 f_2+a_2 a_3 f_2 f_3 +a_2 a_3 a_0 f_2 f_3 f_0}, \\[2mm]
 \smallskip
 \displaystyle 
 \bar{f}_3=a_3 a_0 f_0 
           \frac{1+a_1 f_1+a_1 a_2 f_1 f_2 +a_1 a_2 a_3 f_1 f_2 f_3}
                {1+a_3 f_3+a_3 a_0 f_3 f_0 +a_3 a_0 a_1 f_3 f_0 f_1}. 
\end{array}   \label{qP5}
\een

\remark
Contrary to the case where these mappings are considered to be field
operators, we define the composition of mappings as that of functions. 
For example, 
for $\vp:(x,y)\mapsto(x^2,y)$ and $\psi:(x,y)\mapsto(x+y,y)$, we have
$\psi\circ \vp:(x,y)\mapsto (x^2+y,y)$.\\

By the change of variables
\ben
a_3=1/(a_0 a_1 a_2 q),
&&f_0=x,~ f_1=y,~ f_2=c/x,~ f_3=d/y
\een
this equation reduces to the following 
2-dimensional non-autonomous discrete dynamical
system:
\ben
\varphi: (x,y)&\mapsto& (\ol{x},\ol{y}),
\een
\be \label{qp6}  \left\{ 
\ba{l}
\displaystyle{ \ol{x}= \frac{q a_0a_1 xy + c(d + a_0 d x + q a_0a_1a_2 y)}{
      qx(1 + a_0(x +  a_1a_2 c y + a_1 x y))}}\\[2mm]
\displaystyle{\ol{y}= \frac{c(d + a_0dx + a_0a_1dxy + qa_0a_1a_2 y )}{
       y(q a_0 x(1 + a_1y) + c(d + q a_0a_1a_2 y))}},
\ea \right.
\ee
where the change in the parameters is given by
\ben
&& (\ol{a_0},\ol{a_1},\ol{a_2},\ol{q},\ol{c},\ol{d}) 
= (a_0,a_1,a_2,q,\frac{d}{q},\frac{c}{q}).
\een

\section{Space of initial conditions and Cremona isometries}

\subsection{Space of initial conditions}
 
The notion of space of initial conditions (values) was 
first proposed by Okamoto \cite{okamoto} for the continuous 
Painlev\'e equations and was recently 
applied by Sakai \cite{sakai} for the discrete Painlev\'e equations.
In the discrete case it is linked to automorphisms
of certain families of rational surfaces. 
The relations of surfaces and groups of 
these automorphisms were also studied by many authors 
from the algebraic-geometric view point \cite{do,looijenga}.
In this section, following Sakai's method, we construct the space 
of initial conditions for $\vp$.

Let $X$ and $Y$ be rational surfaces and
let $X'$ and $Y'$ be surfaces obtained by the successive blow-ups 
$\pi_X: X' \to X$ and $\pi_Y: Y' \to Y$.
A rational mapping $\vp': X'\to Y'$ is called {\it lifted}
from a rational mapping $\vp:X\to Y$, if 
$\pi_Y \circ \vp'= \vp \circ \pi_X$ holds for any point 
where $\vp \circ \pi_X$ and $\vp'$ are defined.\\ 

\definition (space of initial conditions). 
Let $X_i$'s be rational surfaces and
let $\{\varphi_i:X_i \to X_{i+1}\}$ be a sequence of birational mappings.
A sequence of rational surfaces $\{Y_{i}\}$ 
is (or $Y_i$ themselves are) called 
the space of initial conditions for $\{\varphi_i\}$ 
if each $\varphi_i$ is 
lifted to an isomorphism, 
i.e. bi-holomorphic mapping, from $Y_i$ to $Y_{i+1}$.\\

\begin{figure}[ht]
\begin{center}
\includegraphics*[]{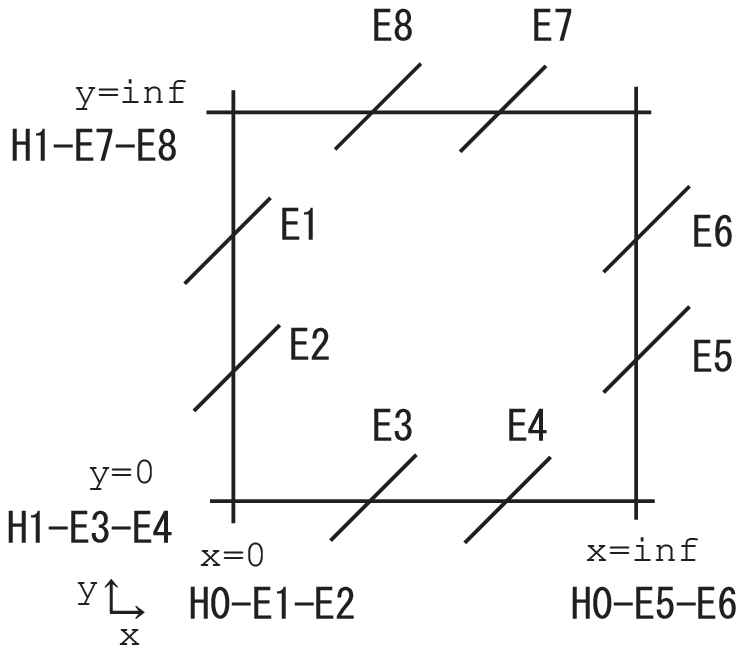}
\caption[]{Irreducible divisors and their intersections}\label{a3}
\end{center}
\end{figure}

Let us consider the mapping (\ref{qp6}) to be a birational
mapping from $\ponet$ to itself. Blowing up successively at the indeterminate
points of $\vp$ and $\vp^{-1}$ (as usual),
we have the surface $X_i$ as described in Fig.\ref{a3}. 
The total transforms of the points of successive blow-ups are 
\begin{eqnarray}\label{blowup}
\ba{ll}
E_1: (x,y)=(0,-d a_3)&
E_2: (x,y)=(0,-a_1)\\
E_3: (x,y)=(-1/a_0,0)&
E_4: (x,y)=(-c/a_2,0)\\
E_5: (1/x,y)=(0,-1/a_1)&
E_6: (1/x,y)=(0,-d/a_3)\\
E_7: (x,1/y)=(-c a_2,0)&
E_8: (x,1/y)=(-a_0,0),
\ea
\end{eqnarray}
where we use $a_3=1/(a_0 a_1 a_2 q)$ instead of $q$.

\begin{thm}
The mapping 
$\vp:\ponet \to \ponet$ (\ref{qp6}) can be lifted to the sequence of 
isomorphisms $\{\vp: X_i \to X_{i+1}\}$.
\end{thm}

Let $X$ be a rational surface which is given by blow-ups (\ref{blowup})
from $\ponet$, where $a_i,c,d\in \mc \setminus \{0\}$. 
We denote the linear equivalence class of total transforms of
$x={\rm constant},$ ($y={\rm constant}$) on
$X$ by $H_0$ (resp. $H_1$).
The Picard group of the surface $X$, i.e. the group of 
linear equivalence classes of 
divisors on $X$, is denoted as $\pic(X)$.
The mapping $\vp$ induces the push-forward
mapping $\vp_*:\pic(X_i) \to \pic(X_{i+1})$ ($\pic(X_i) \simeq \pic(X_{i+1})$)
as
\ben \ba{ll}
H_0 \mapsto 2H_0+H_1-E_2-E_4-E_6-E_8&\\
H_1 \mapsto H_0+2H_1-E_2-E_4-E_6-E_8&\\
E_1 \mapsto H_0+H_1-E_2-E_4-E_8 &
E_2 \mapsto E_5 \\
E_3 \mapsto H_0+H_1-E_2-E_4-E_6&
E_4 \mapsto E_7\\
E_5 \mapsto H_0+H_1-E_4-E_6-E_8&
E_6 \mapsto E_1\\
E_7 \mapsto H_0+H_1-E_2-E_6-E_8&
E_8 \mapsto E_3~.\\
\ea \een

\subsection{The group of Cremona isometries}

In this section we shall consider automorphisms of the family 
of $X$ with various parameters.

The Picard group of $X$ is the ${\mathbb Z}$-module (lattice)
$${\rm Pic}(X) = 
{\mathbb Z}H_0 + {\mathbb Z}H_1+  {\mathbb Z}E_1+\cdots+{\mathbb Z}E_8.$$
The intersection number of any two divisors can be calculated
by using the intersection form
\begin{eqnarray}\label{isn}
H_i \cdot H_j = 1-\delta_{i,j},~ E_k\cdot E_l= -\delta_{k,l},~
H_i \cdot E_k=0~.
\end{eqnarray}

Let $D_0,D_1,D_2,D_3$ be divisors as
\ben
\begin{array}{ll}
D_0=H_0-E_1-E_2,& D_1=H_1-E_3-E_4\\
D_2=H_0-E_5-E_6,& D_3=H_1-E_7-E_8.
\end{array}
\een
The anti canonical divisor of $X$, 
$-K_{X}=2H_0 + 2H_1- E_1-\cdots-E_8$, is uniquely 
decomposed into prime divisors (for generic parameters) as  
$$ -K_X= D_0+D_1+D_2+D_3.$$
The connections of the $D_i$ are expressed by the following diagram.
\begin{eqnarray}\label{ddynkin}
\begin{picture}(0,45)(50,10)
\put(-2.5,20){\circle{5}}
\put(-2.5,10){\makebox(0,0){$D_1$}}
\put(42.5,20){\circle{5}}
\put(42.5,10){\makebox(0,0){$D_2$}}
\put(87.5,20){\circle{5}}
\put(87.5,10){\makebox(0,0){$D_3$}}
\put(42.5,42.5){\circle{5}}
\put(42.5,52.5){\makebox(0,0){$D_0$}}
\put(0,20){\line(1,0){40}}
\put(45,20){\line(1,0){40}}
\put(-0.5,22.0){\line(2,1){40}}
\put(85.5,22.0){\line(-2,1){40}}
\end{picture}
\end{eqnarray}
Such rational surfaces are said to be of the $A_3^{(1)}$ type which is  
the type of those surfaces in Sakai's list associated with the $q$-$P_{VI}$ 
equation .\\

\definition. 
An automorphism $s$ of $\pic(X)$ is called a {\it Cremona isometry}
\cite{do,looijenga}
if $s$ preserves 
i) the intersection form on $\pic(X)$,
ii) the canonical divisor $K_X$, 
iii) the semigroup of effective classes of divisors. \\

Let $X$ and $X'$ be rational surfaces.
In general, if a mapping $\vp$ is an isomorphism from
$X$ to $X'$, its action on the Picard group $\vp_*:\pic(X) \to \pic(X')$
(or $\vp^*:\pic(X')\to\pic(X)$) ($\pic(X) \simeq \pic(X')$)is always a Cremona isometry.\\

A Cremona isometry preserves the set $\{D_i\}$ and its orthogonal
(with respect to the intersection form) lattice,
\be \label{Q} 
&&Q:=
{\mathbb Z}\alpha_0 \oplus{\mathbb Z}\alpha_1\oplus{\mathbb Z}\alpha_2\oplus
{\mathbb Z}\alpha_3\oplus {\mathbb Z}\alpha_4 \oplus {\mathbb Z} \alpha_5
\ee
where the $\alpha_i$'s are
\ben
\begin{array}{ll}
\alpha_0=E_5-E_6,& \alpha_1=E_1-E_2\\
\alpha_2=H_1-E_1-E_5,& \alpha_3=H_0-E_3-E_7\\
\alpha_4=E_3-E_4,& \alpha_5=E_7-E_8~.\\
\end{array}
\een

We define the action $w_i$ on $\pic(X)$ corresponding to $\alpha_i$
as 
$$ w_i(D)= D - \frac{2 D\cdot \alpha_i}{\alpha_i \cdot \alpha_i}\alpha_i
$$
where $D\in \pic(X)$.
The Cartan matrix is as follows.
\ben
(c_{i,j})&=&
 \left(
  \begin{array}{rrrrrr}
    2 &  0 & -1 &  0 &  0&  0\\
    0 &  2 & -1 &  0 &  0&  0\\
   -1 & -1 &  2 & -1 &  0&  0\\
    0 &  0 & -1 &  2 & -1& -1\\
    0 &  0 &  0 & -1 &  2&  0\\
    0 &  0 &  0 & -1 &  0&  2
  \end{array}
 \right). 
\begin{picture}(200,30)(50,30)
\put(100,20){\line(1,0){20}}
\put(125,20){\line(1,0){20}}
\put(150,20){\line(1,0){20}}
\put(122.5,22.5){\line(0,1){20}}
\put(147.5,22.5){\line(0,1){20}}
\put(97.5,20){\circle{5}}
\put(122.5,20){\circle{5}}
\put(147.5,20){\circle{5}}
\put(172.5,20){\circle{5}}
\put(122.5,45){\circle{5}}
\put(147.5,45){\circle{5}}
\put(97.5,10){\makebox(0,0){$\alpha_{1}$}}
\put(122.5,10){\makebox(0,0){$\alpha_{2}$}}
\put(147.5,10){\makebox(0,0){$\alpha_{3}$}}
\put(172.5,10){\makebox(0,0){$\alpha_{4}$}}
\put(135,50){\makebox(0,0){$\alpha_{0}$}}
\put(160,50){\makebox(0,0){$\alpha_{5}$}}
\end{picture}
\een

We also define the actions of the generators of automorphisms of the
Dynkin diagrams $\sigma_{102345}$ and $\sigma_{543210}$ as
\be
\sigma_{102345}:&& (H_0,H_1,E_1,E_2,E_3,E_4,E_5,E_6,E_7,E_8) \nonumber  \\
&& \mapsto (H_0,H_1,E_5,E_6,E_3,E_4,E_1,E_2,E_7,E_8) \nonumber  \\
\sigma_{543210}:&& (H_0,H_1,E_1,E_2,E_3,E_4,E_5,E_6,E_7,E_8) \nonumber \\
&& \mapsto (H_1,H_0,E_3,E_4,E_1,E_2,E_7,E_8,E_5,E_6),
\ee
where the suffixes of $\sigma$ mean corresponding permutations.
For convenience we also use the automorphism 
$\sigma_{012354}=\sigma_{543210} \circ \sigma_{102345} \circ \sigma_{543210}$.
For simplicity we write $\sigma_{102345}$ and
$\sigma_{012354}$ as $\sigma_{10}$ and
$\sigma_{54}$ respectively.

Here, the group of automorphisms of the lattice $\bigoplus {\mathbb Z}\alpha_i$
which preserve the intersection form (considered to be the inner product) is
$\pm \wt{W}(D_5^{(1)})$ (see \cite{kac} \S 5.10). Since each element of 
$-\wt{W}(D_5^{(1)})$ does not preserve effectiveness of divisor classes, 
we have the following theorem (it can be confirmed directly 
that $\wt{W}(D_5^{(1)})$ does preserve effectiveness 
by realization as automorphisms of the family of surfaces).

\begin{thm} \label{cremona}
The group of Cremona isometries of the surface $X$ is 
$\wt{W}(D_5^{(1)})$.
\end{thm}

\begin{cor}
The mapping $\varphi^2$ acts on the root basis as 
\ben
&&\varphi^2:(\alpha_0,\alpha_1,\alpha_2,\alpha_3,\alpha_4,\alpha_5)
\mapsto \\&&~(\alpha_0+K,\alpha_1+K,\alpha_2-K,
   \alpha_3-K,\alpha_4+K,\alpha_5+K)
\een
where $K=-K_X=\alpha_0+\alpha_1+2\alpha_2+2\alpha_3+\alpha_4+\alpha_5$
corresponds to the canonical central element
and $\varphi$ is written by generators as
\ben
\vp&=&\sigma_{54}\circ \sigma_{10}\circ w_5\circ w_4\circ 
    w_1\circ w_0\circ w_2\circ w_3\circ w_2 .
\een
\end{cor}

\remark
Since we have defined the composition of mappings as in Sect.~2,
the composition of actions on the Picard group should like
a changes of basis, i.e. for $\vp:(x,y)\mapsto(2x ,y)$ and 
$\psi:(x,y)\mapsto(x+y,y)$, we have
$\psi\circ \vp:(x,y)\mapsto (2x +2y,y)$.\\

\subsection{Realization of Cremona isometries}

Each Cremona isometry of Th.\ref{cremona} can be realized as birational
mapping on $\ponet$ which is lifted to an automorphism of 
the family of surfaces $X$ (cf. \cite{sakai,takenawa}).

The birational actions of generators are as follows:
$(x,y,a_0,a_1,a_2,a_3,c,d)$ is mapped to 
\ben
&
\left(x, y \sqrt{\frac{a_3}{a_1 d}};~  
a_0, \sqrt{\frac{a_1 a_3}{d}}, a_2, 
\sqrt{a_1 a_3 d},c,\frac{a_3}{a_1}\right)&\mbox{by $w_0$  }\\
&
\left(x, y\sqrt{\frac{a_1}{a_3 d}};~  
a_0, \sqrt{a_1 a_3 d}, a_2, 
\sqrt{\frac{a_1 a_3}{d}}, c, \frac{a_1}{a_3}\right)&\mbox{by $w_1$  } \\
&
\left(\frac{x (a_1 y+1)}{(y+a_3 d)} \sqrt{\frac{a_3 d}{a_1}},
\frac{y}{\sqrt{a_1 a_3 d}};~ 
a_0 \sqrt{a_1 a_3 d}, \sqrt{\frac{a_1}{a_3 d}}, 
a_2 \sqrt{a_1 a_3 d}, \sqrt{\frac{a_3}{a_1 d}}, 
c, \frac{1}{a_1 a_3}\right)&\mbox{by $w_2$  }\\
&
\left(\frac{x}{\sqrt{a_0 a_2 c}},
\frac{y (x+ a_2 c)}{(a_0 x+ 1)}\sqrt{\frac{a_0}{a_2 c}};~
\sqrt{\frac{a_0}{a_2 c}}, a_1 \sqrt{a_0 a_2 c}, 
\sqrt{\frac{a_2}{a_0 c}}, a_3 \sqrt{a_0 a_2 c}, 
\frac{1}{a_0 a_2}, d\right)&\mbox{by $w_3$  }\\
&
\left(x \sqrt{\frac{a_2}{a_0 c}}, y;~ 
\sqrt{\frac{a_0 a_2}{c}}, a_1, \sqrt{a_0 a_2 c}, 
a_3, \frac{a_2}{a_0}, d\right)&\mbox{by $w_4$  }\\
&
\left(x \sqrt{\frac{a_0}{a_2 c}}, y;~ 
\sqrt{a_0 a_2 c}, a_1, \sqrt{\frac{a_0 a_2}{c}}, 
a_3, \frac{a_0}{a_2}, d\right)&\mbox{by $w_5$  }\\
&
\left(\frac{1}{x},\frac{y}{d};~ \frac{1}{a_0}, \frac{1}{a_3},
\frac{1}{a_2}, \frac{1}{a_1}, \frac{1}{c}, \frac{1}{d}\right)&\mbox{by $\sigma_{10}$  }\\
&
\left(\frac{y}{d}, \frac{x}{c};~ \frac{1}{a_3}, \frac{1}{a_2},
\frac{1}{a_1}, \frac{1}{a_0}, \frac{1}{d}, \frac{1}{c}\right)&\mbox{by $\sigma_{543210}$  }.
\een

\remark
If one would prefer to relate these transformation to Sakai's paper 
\cite{sakai} explicitly,
it is sufficient to put $f,g$ and $b_1,b_2,\cdots,b_8$ as
$f= x, g= y$ and
$b_1=-a_1, b_2=-da_3, b_3=-1/a_1, b_4=-d/a_3,
b_5=-1/a_0, b_6=-c/a_2, b_7=-a_0, b_8=-ca_2$.
(This coincidence is up to the coefficients of $f$ and $g$ in each
birational transformation, because the normalization
is different in Sakai's paper. While 
there are 8 parameters in that paper, there are only 6 parameters here.
If we consider surfaces whose Picard group and effective 
classes are the same as those of the surfaces $X$ and
normalize by automorphisms of $\ponet$, then
only ``6'' parameters remain.) 

\subsection{Relation to the $q$-Painlev\'e VI equation}

We define the mapping $\psi$ as 
$\psi=\sigma_{10} \circ \sigma_{543210} \circ w_3 \circ w_5 \circ w_4 \circ 
w_3$, then $\psi$ is expressed as
\ben
\psi: (x,y)&\mapsto& (\ol{x},\ol{y}),
\een
\be \left\{ \ba{l}
\displaystyle{\ol{x}= \frac{d(1+a_0x)(c+a_2x)}{y(x+a_0)(x+ca_2)}} \\
\displaystyle{\ol{y}= x}
\ea \right. 
\ee
with the change of parameters:
\ben
&&(\ol{a_0},\ol{a_1},\ol{a_2},\ol{q},\ol{c},\ol{d}) = 
(\frac{a_0}{q}, q a_1,\frac{a_2}{q}, q, d, c).
\een
The mapping $\psi^2$ is the $q$-$P_{VI}$ equation.
Actually it acts on the root basis as 
\ben
&&(\al_0,\al_1,\al_2,\al_3,\al_4,\al_5)
\mapsto (\al_0,\al_1,\al_2+K,\al_3-K,\al_4,\al_5).
\een

\remark
If one would prefer to relate these transformation to Sakai's $qP_{VI}$ 
\cite{sakai},
it is sufficient to put 
$f= x, g= y$ and
$b_1=-t a_1, b_2=-t da_3, b_3=-t/a_1, b_4=-t d/a_3,
b_5=- t/a_0, b_6=-t c/a_2, b_7=-ta_0, b_8=-tca_2$
and $\ol{t}=qt$ and $q'=q^2$ (the normalization is changed by 
introducing the new variable $t$).

\begin{prop}
The mapping $\vp^2$ can be decomposed into conjugate mappings of 
$q$-$P_{VI}(=\psi^2)$ as 
\ben
&&\varphi^2=(\sigma_{543210}\circ w_2\circ \mbox{$q$-$P_{VI}$} \circ w_2\circ 
\sigma_{543210})  \circ (w_2\circ \mbox{$q$-$P_{VI}$} \circ w_2).
\een

\end{prop}

\section{The group of commutative elements}\label{comm}

In this section we investigate the subgroup of 
$\wt{W}(D_5^{(1)})$ whose elements commute with $\varphi^2$
(we denote it as $W_C$).
We prove the following theorem.

\begin{thm}
$$W_C \simeq \mz \times \wt{W}(A_3^{(1)}) \times \wt{W}(A_1^{(1)})$$
holds, where the meaning of each symbol is: 

\ni
{\rm i)} $\mz$ is $\vp^{2\mz}$;

\ni {\rm ii)} $\wt{W}(A_3^{(1)})$ 
is the extended Weyl group defined by the root basis 
$$(\beta_0,\beta_1,\beta_2,\beta_3):=
(\al_3+\al_5,~ \al_1+\al_2,~ \al_3+\al_4,~ \al_0+\al_2)$$
and the automorphisms of Dynkin diagram 
$\sigma_{\beta(1032)}$ and $\sigma_{\beta(3210)}$;
It coincides with the original $W(A_3^{(1)})$ except for the extension;

\ni {\rm iii)} $\wt{W}(A_1^{(1)})$
is the extended Weyl group defined by the root basis 
$$(\gamma_0,\gamma_1):=(\al_0+\al_1+\al_2+\al_3,~ \al_2+\al_3+\al_4+\al_5)$$
and the automorphisms of Dynkin diagram $\sigma_{\gamma(10)}$.
\end{thm}

Notice that the subset of the lattice $Q$ (\ref{Q}) whose elements are 
preserved  by $\vp^2$ is
$$Q_B:=\{\sum_{i=0}^5 a_i\al_i \in Q;~
 a_0+a_1-a_2-a_3+a_4+a_5=0\}.$$
The lattice $Q_B$ can be described as
\be \al \in  Q_B&=&{\rm Span}_\mz (\beta_0,\beta_1,\beta_2,\beta_3,\gamma_0)
\nonumber \\
&=& {\rm Span}_\mz (\beta_0,\beta_1,\beta_2,\gamma_0,\gamma_1).
\ee

\begin{lemma}\label{lemma1}
The group $W_C$ preserves the lattice $Q_B$.
\end{lemma}

\bp
Let $w_c \in W_C$  and let $q\in Q_B$. Since 
$\vp^2 \circ w_c(q) = w_c\circ \vp^2(q) = w_c(q)$,  $w_c(q)$ is preserved by
$\vp^2$ and therefore $w_c(q) \in Q_B$.
\ep

\begin{lemma}\label{lemma2} 
The group whose actions preserve the lattice $Q_B$ and the intersection form
is $\pm\wt{W}(A_3^{(1)}) \times \pm\wt{W}(A_1^{(1)})$.
\end{lemma}

\bp
First we consider $w(\beta_i)$.
$w(\beta_i)$ is written as 
$$w(\beta_i)=b_{i,0}\beta_0+b_{i,1}\beta_1+b_{i,2}\beta_2+b_{i,3}\beta_3
+c_{i,0}\gamma_0,$$
where $b_{i,j},c_{i,j}\in \mz$. 
From the equation 
\ben
-2 &=& \beta_i\cdot \beta_i= w(\beta_i)\cdot w(\beta_i)\\ 
&=&-(b_{i,0}-b_{i,1})^2-(b_{i,1}-b_{i,2})^2-(b_{i,2}-b_{i,3})^2-
(b_{i,3}-b_{i,0})^2-2(c_{i,0})^2,
\een
we have $b_{i,0}=b_{i,1}=b_{i,2}=b_{i,3}$ or $c_{i,0}=0$.
In the former case we have that
$$1=w(\beta_i) \cdot w(\beta_{i+1})= - 2 c_{i,0} c_{i+1,0},$$
which is a contradiction.
Hence $w$ preserves the lattice $\mz\beta_0+\mz\beta_1+\mz\beta_2+\mz\beta_3$
and therefore the action of $w$ on this lattice
coincides with an element of the extended affine Weyl group 
$\pm\wt{W}(A_3^{(1)})$ (See \cite{kac} \S 5.10).

Next we consider $w(\gamma_0)$ and $w(\gamma_1)$.
$w(\gamma_i)$ is written as 
$$w(\gamma_i)=b_{i,0}\beta_0+b_{i,1}\beta_1+b_{i,2}\beta_2+b_{i,3}\beta_3
+c_{i,0}\gamma_0$$
and we have $b_{i,0}=b_{i,1}=b_{i,2}=b_{i,3}$ or $c_{i,0}=0$ for each $i$.
Since the rank of $w(Q_B)$ equals that of $Q_B$, 
$c_{i,0}$ can not be zero.
Hence we have 
$w(\gamma_i)\in \mz K \pm \gamma_0\subset \mz \gamma_0+\mz\gamma_1$
and therefore the action of $w$ on the lattice $\mz \gamma_0+\mz\gamma_1$
coincides with an element of the extended affine Weyl group 
$\pm\wt{W}(A_1^{(1)})$.
\ep\\

Now we are ready to prove the theorem.\\

\bp
By Lemma~\ref{lemma1} and Lemma~\ref{lemma2} 
it follows that the action of $W_C \subset \wt{W}(D_5^{(1)})$ on $Q_B$
is given by $\wt{W}(A_3^{(1)}) \times \wt{W}(A_1^{(1)})$.
One of the extensions of this action onto $Q$ is given 
in appendix~\ref{app2}.

We investigate the variety of extensions.
Suppose the actions of $s_1, s_2 \in W_C$ on the lattice $Q_B$ are the same.
We write $s_1\circ s_2^{-1}(x)$ as $\ol{x}$.   
Since $\al_0,\beta_0,\beta_1,\beta_2,\beta_3,\gamma_0$ are linearly
independent, $\ol{\al_0}$ can be written as
$$a_0 \al_0+b_0\beta_0+b_1\beta_1+b_2\beta_2+b_3\beta_3+c_0\gamma_0.$$
Using the relations 
$\al_0\cdot \beta_i=\ol{\al_0}\cdot \ol{\beta_i}= \ol{\al_0}\cdot \beta_i$
and
$\al_0\cdot \gamma_0=\ol{\al_0}\cdot \ol{\gamma_0}= \ol{\al_0}\cdot \gamma_0$,
we have 
\ben
\left\{
\ba{ccc}
-2b_0+b_1+b_3&=&0\\
a_0+b_0-2b_1+b_2&=&1\\
b_1-2b_2+b_3&=&0\\
-a_0+b_0+b_2-2b_3&=&-1\\
-a_0-2c_0&=&-1 .\\
\ea
\right. 
\een
and hence 
\ben 
\ol{\al_0}&=&\al_0+\mz K+\mz(2\al_0+\beta_0+2\beta_1+\beta_2-\gamma_0)\\
&=&\al_0+\mz K+\mz (\al_2+\al_3).
\een
Moreover from $(\ol{\al_0})^2=(\al_0)^2=-2$, we have
\be
\ol{\al_0}&=&\al_0+\mz K \mbox{ or } \al_0+\al_2+\al_3+\mz K.
\ee
Since $\ol{\beta_i}=\beta_i$ and $\ol{\gamma_0}=\gamma_0$,
we have: if $\ol{\al_0}=\al_0 + z K,$ ($z\in \mz$), then
\ben
&&(\ol{\al_0},\ol{\al_1},\ol{\al_2},\ol{\al_3},\ol{\al_4},\ol{\al_5})\\
&=&(\al_0 + z K, \al_1 + z K, \al_2 - z K, al_3 - z K, 
\al_4 + z K, \al_5 + z K)
\een
and if $\ol{\al_0}=\al_0+\al_2+\al_3+ z K,$ ($z\in \mz$), then
\ben
&&(\ol{\al_0},\ol{\al_1},\ol{\al_2},\ol{\al_3},\ol{\al_4},\ol{\al_5})\\
&=&(\al_0 +\al_2+\al_3 + z K, \al_1 +\al_2+\al_3+ z K, -\al_3 - z K,\\
&&-\al_2 - z K, \al_2+\al_3+ \al_4 + z K, \al_2+\al_3+\al_5 + z K).
\een
In the former case $s_1\circ s_2^{-1} =\vp^2 = (r_1 \circ r_0)^z$ and
in the later case  $s_1\circ s_2^{-1} =\vp^2 = r_0 \circ (r_1 \circ r_0)^z$.
Since $r_0$ does not commute with $\vp^2$ 
($r_0 \circ \vp^2 = \vp^{-2} \circ r_0$),  
the action of $W_B$ is extended by $\vp^{2\mz}$. 
\ep\\

{\noindent{\it Acknowledgment.}}
The author would like to thank H. Sakai, J. Satsuma and R. Willox 
for discussions and advice. The author is also grateful
to the referee for his useful comments and suggestions.

\appendix

\section{Embedding $\wt{W}(A_1^{(1)}\times A_3^{(1)})$ of KNY 
into $\wt{W}(D_5^{(1)})$}

By studying the actions of the elements of original Weyl group 
$\wt{W}(A_1^{(1)}\times A_3^{(1)})$ on the root basis,
we can realize its embedding into the $\wt{W}(D_5^{(1)})$
as follows
\ben
r_0 &=& w_2\circ w_3 \circ w_2\\
r_1 &=& w_0\circ w_1 \circ w_4 \circ w_5 \circ w_2 \circ w_3 \circ w_2
 \circ w_0\circ w_1 \circ w_4 \circ w_5\\
\pi &=& w_0 \circ w_1 \circ w_4 \circ w_5 \circ \sigma_{10} \circ \sigma_{54}\\
s_0 &=& w_5 \circ w_3 \circ w_5\\
s_1 &=& w_1 \circ w_2 \circ w_1\\
s_2 &=& w_3 \circ w_4 \circ w_3\\
s_3 &=& w_0 \circ w_2 \circ w_0\\
\rho &=& \sigma_{543210} \circ \sigma_{10}.
\een

\section{Embedding the group 
$W_C$
into $\wt{W}(D_5^{(1)})$} \label{app2}

As mentioned in Section~\ref{comm} the actions of 
$\wt{W}(A_3^{(1)}) \times \wt{W}(A_1^{(1)})$ on the lattice $Q_B$
are not uniquely expanded to onto the lattice $Q$. The variety of
extensions is $\vp^{2\mz}=(r_1\circ r_0)^{\mz}$.
Here, we give one of the embeddings of 
$\wt{W}(A_3^{(1)}) \times \wt{W}(A_1^{(1)})$
into $\wt{W}(D_5^{(1)})$.

For $\beta=\beta_i$ or $\gamma_i$ we define the action on the lattice $Q$
as 
$$w_{\beta}(\al)=\al-2\frac{\beta\cdot\al}{\beta\cdot\beta}\beta,$$ 
where $\al\in Q$.
In order to find its description in terms of the fundamental elements 
of $\wt{W}(D_5^{(1)})$,
we use the relation
\ben
w_{w_{\al}(\beta)}=w_{\al}\circ w_{\beta}\circ w_{\al},
\een
where $\al$ and $\beta$ are real roots.
\ben
w_{\beta_0} &=& w_{\al_3+\al_5}=w_3\circ w_5\circ w_3\\
w_{\beta_1} &=& w_{\al_1+\al_2}=w_1\circ w_2\circ w_1\\
w_{\beta_2} &=& w_{\al_3+\al_4}=w_3\circ w_4\circ w_3\\
w_{\beta_3} &=& w_{\al_0+\al_2}=w_0\circ w_2\circ w_0\\
w_{\gamma_0} &=& w_{\al_0+\al_1+\al_2+\al_3}=
   w_3\circ w_0\circ w_2\circ w_1\circ w_2\circ w_0\circ w_3\\
w_{\gamma_1} &=& w_{\al_2+\al_3+\al_4+\al_5} =
   w_2\circ w_5\circ w_3\circ w_4\circ w_3\circ w_5\circ w_2\\
\een
Finally, we define the action of automorphisms of the Dynkin diagram
of $\wt{W}(A_3^{(1)}) \times \wt{W}(A_1^{(1)})$: $\sigma_{\beta(1032)},
\sigma_{\beta(1032)},\sigma_{\gamma(10)}$ as 
$(\al_0,\al_1,\al_2,\al_3,\al_4,\al_5)$ is mapped to 
\ben
\mbox{by } \sigma_{\beta(1032)}: &&
(\al_0+\al_1+\al_2+\al_3+\al_4, \al_0+\al_1+\al_2+\al_3+\al_5,
-\al_0-\al_1-\al_2, \\&& -\al_3-\al_4-\al_5,
\al_0+\al_2+\al_3+\al_4+\al_5, \al_1+\al_2+\al_3+\al_4+\al_5)\\
\mbox{by } \sigma_{\beta(3210)}:&&
(\al_0+\al_1+\al_2+\al_3+\al_5, \al_0+\al_1+\al_2+\al_3+\al_4,
-\al_0-\al_1-\al_2,\\&& -\al_3-\al_4-\al_5,
\al_1+\al_2+\al_3+\al_4+\al_5, \al_0+\al_2+\al_3+\al_4+\al_5)\\
\mbox{by } \sigma_{\gamma(10)}:&&
(\al_0+\al_2+\al_3+\al_4+\al_5, \al_1+\al_2+\al_3+\al_4+\al_5,
-\al_3-\al_4-\al_5,\\&& -\al_0-\al_1-\al_2,
\al_0+\al_1+\al_2+\al_3+\al_4, \al_0+\al_1+\al_2+\al_3+\al_5).
\een
Hence, these are written in termes of 
the fundamental elements of $\wt{W}(D_5^{(1)})$ as: 
\ben
\sigma_{\beta(1032)}&=&\sigma_{543210}
\circ w_5\circ w_4 \circ w_1\circ w_0\circ w_2\circ w_3\circ w_2 \\
\sigma_{\beta(3210)}&=&\sigma_{543210}\circ \sigma_{54}\circ \sigma_{10}
\circ w_5\circ w_4 \circ w_1\circ w_0\circ w_2\circ w_3\circ w_2 \\
\sigma_{\gamma(10)}&=&\vp = \sigma_{54}\circ \sigma_{10}
\circ w_5\circ w_4 \circ w_1\circ w_0\circ w_2\circ w_3\circ w_2 .
\een

\end{document}